\documentclass[a4paper,11pt]{article}

\usepackage{graphicx}
\usepackage{amsmath}
\usepackage{amssymb}
\usepackage{authblk}
\usepackage{fancyhdr}
\usepackage[titletoc]{appendix}
\usepackage{multicol}
\usepackage{subcaption}
\usepackage{bm}
\usepackage{amsfonts}
\usepackage{epstopdf}
\usepackage{xcolor}
\usepackage{color}
\usepackage{wasysym}
\usepackage[colorlinks=true, allcolors=blue]{hyperref}

\usepackage{hhline}

\newcommand{\rc}[1]{\textcolor{black}{#1}}

\title{\sffamily{Twisted plasma waves driven by twisted ponderomotive force}}

\date{\today}

\author[1]{Y. Shi}
\author[2]{D. R. Blackman}
\author[3]{R. J. Kingham}
\author[2]{A. V. Arefiev}
\affil[1]{Department of Plasma Physics and Fusion Engineering, University of Science and Technology of China, Hefei 230026, China}%
\affil[2]{Department of Mechanical and Aerospace Engineering, University of California at San Diego, La Jolla, CA 92093, USA}
\affil[3]{Blackett Laboratory, Imperial College London, London SW7 2AZ, United Kingdom}

\date{\today}

\begin{document}

\vskip -2.0cm
\maketitle
\vskip -3.0cm

\begin{abstract}
We present results of twisted plasma waves driven by twisted ponderomotive force.
With beating of two, co-propagating, Laguerre-Gaussian (LG) orbital angular momentum (OAM) laser pulses with different frequencies and also different twist indices, we can get twisted ponderomotive force.  Three-dimensional particle-in-cell simulations are used to demonstrate the twisted plasma waves driven by lasers. The twisted plasma waves have an electron density perturbation with a helical rotating structure. Different from the predictions of the linear fluid theory, the simulation results show a nonlinear rotating current and a static axial magnetic field.  Along with the rotating current is the axial OAM carried by particles in the twisted plasma waves.  Detailed theoretical analysis of twisted plasma waves is given too.
\end{abstract}

\section{\sffamily{Introduction}}
High-power laser interactions with plasma have been widely studied since the 1970s. Most of these studies and applications are related to energy and momentum coupling from laser to plasma. Applications include particle accelerators~\cite{Esarey2009}, X-ray sources, inertial confinement fusion, etc. Plasma wave is an important collective effect driven by high-power laser pulses. Energy and momentum are proven to be exchanged in high efficiency between charged particles and field (laser or plasma field) in this process, which are the foundations of many applications. For example, the self-generated magnetic field plays an important role in laser-plasma interactions. The Inverse Faraday (IF) effect for circularly polarized light~\cite{Ali2010, Haines2001, Najmudin2001, Sheng1996} is one of the well-known methods of laser-driven {direct current} magnetic field generation. {It can be explained as} the results of spin angular momentum (SAM) transfer from light to plasma. For a circularly polarized light beam, every photon has a SAM of $\hbar$. On the other hand, light can also possess orbital angular momentum (OAM)~\cite{Allen1992} and thereby have the potential to create plasmas with OAM. The OAM of a photon is due to the helical wavefront instead of the circular polarization. Mathematically, {we can use}  a basis set of orthogonal Laguerre-Gaussian (LG) modes to express any helical wavefronts. For one pure LG mode with a twist index of $l$, every photon has $l\hbar$ of OAM. While such twisted light in conventional optics at low intensities is widely studied(e.g. light tweezers ~\cite{Yao2011}), high intensities ($I > 10^{16}$~W/cm$^2$) twisted light received moderate attention very recently. Unavoidably, the interaction medium will be plasma. Various new simulation phenomena and theories on interactions between intense LG mode laser beams and plasma have been proposed~\cite{shi2014, vieira2016, zhang2016, Zhang2015, vieira2014, wang2015, Zhang2014,vieira2018,Longman2017, Ju2018, Zhu2019,TIKHONCHUK2020,Nuter2018,Blackman2020, Longman2021}. In experiments,  several works have been reported~\cite{Leblanc2017,Denoeud2017,Longman2017,Longman2020, BAE2020, Aboushelbaya2020, Xu2020}. But high-efficiency OAM exchange between plasma and field still needs more attention.

In previous work~\cite{Shi2018},  electron plasma waves with a helical rotating structure and a static, axial magnetic field generation have been confirmed using three-dimensional particle-in-cell (PIC) simulations. Now, we will give a general theory and more details under the condition of paraxial approximation where the spot size of the laser beam is big enough. The ponderomotive force of beating OAM lasers will be analyzed first. Then, the electrons will be driven to oscillate in a way to generate a twisted electron density. The associated axial magnetic field will be calculated using the general theory and compared with the simulation results. The studies on the damping of twisted plasma waves and the axial magnetic field have been published in papers~\cite{Blackman2019a, Blackman2019b, Blackman2020}. This paper is organized as follows: In Section~\ref{sec:PondPotnl}, we demonstrate the ponderomotive potential for two beating LG laser pulses and one LG laser pulse. {Both of them are different from} the ponderomotive potential of Gaussian pulse. In Section~\ref{sec:SimRst}, we show the electric field, the helical electron density and the magnetic field generation of twisted plasma waves in simulations. In Section~\ref{sec:theory}, we give a general theory of twisted plasma waves driven by beating of two LG beams. The helical electron density and the magnetic field are accurately described in theory.  In Section~\ref{sec:LgG}, we explore the scheme of using the beating of a LG beam and a Gaussian beam.


\section{\sffamily{The twisted ponderomotive potential using two beating LG-CP laser pulses}} \label{sec:PondPotnl}
It is convenient to write down a Laguerre-Gaussian (LG) transverse electromagnetic field with polarisation vector $\mathrm{\hat{e_s}}$ in cylindrical coordinates $(r, \theta, x)$ as 
\begin{equation}
  \mathbf{E} = \mathrm{\hat{e_s}}\sum_{p,l}E_{p,l}F_{p,l}(X)\exp\left(ikx -i\omega t +i l\theta -i\phi_{p,l} -\frac{ikw_b^2(x)X^2}{4f(x)}\right) + c.c.
  \label{e_lg}
\end{equation}
with the eigenfunction $F_{p,l}(X)$ given as
  $F_{p,l}(X)=C_{p,l}X^{|l|/2} e^{-X/2}L^{l}_p(X) $
where the normalised radial co-ordinate $X=2r^2/w^2(x)$ is specified, $L_p^{l}(X)$ is a Laguerre polynomial with $C_{pl} = \sqrt{2p!/[\pi (p + |l|)!]}$ ($C_{01} = \sqrt{2/\pi} \simeq 0.8$)  being a normalisation to ensure an orthonormal set, the gouy phase $\phi_{p,l}=(2p+l+1)\arctan(x/x_r)$ and front surface curvature $f(x) = x+x_R^2/x$, the beam width $w_b(x) =w_{b,0}\sqrt{1+x^2/x^2_r}$, the Rayleigh range is $x_r=kw^2_{b,0}$. For the following analysis, we simplify it by only considering the region within the Rayleigh range so that terms with $f(x)$ and $\phi_{p,l}$ can be ignored and we consider $w_b(x)=w_{b,0}$. Now we have an electric field from the laser as $ \mathbf{E} = \mathrm{\hat{e_s}}\sum_{p,l}E_{p,l}F_{p,l}(X)\exp\left(ikx -i\omega t +i l\theta\right)$.

\begin{figure*}
    \centering
    \includegraphics[width=0.4\linewidth]{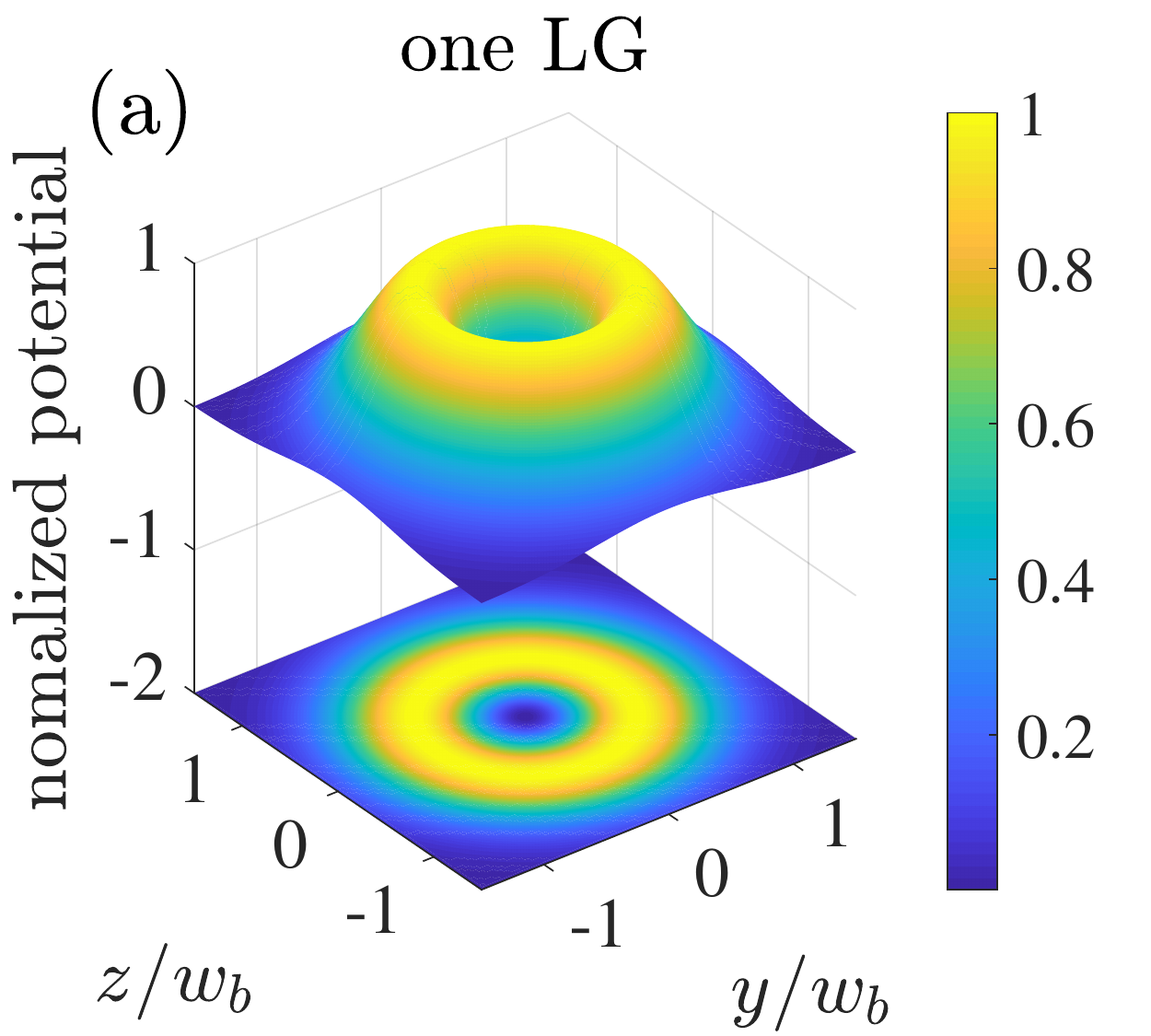}
    \includegraphics[width=0.4\linewidth]{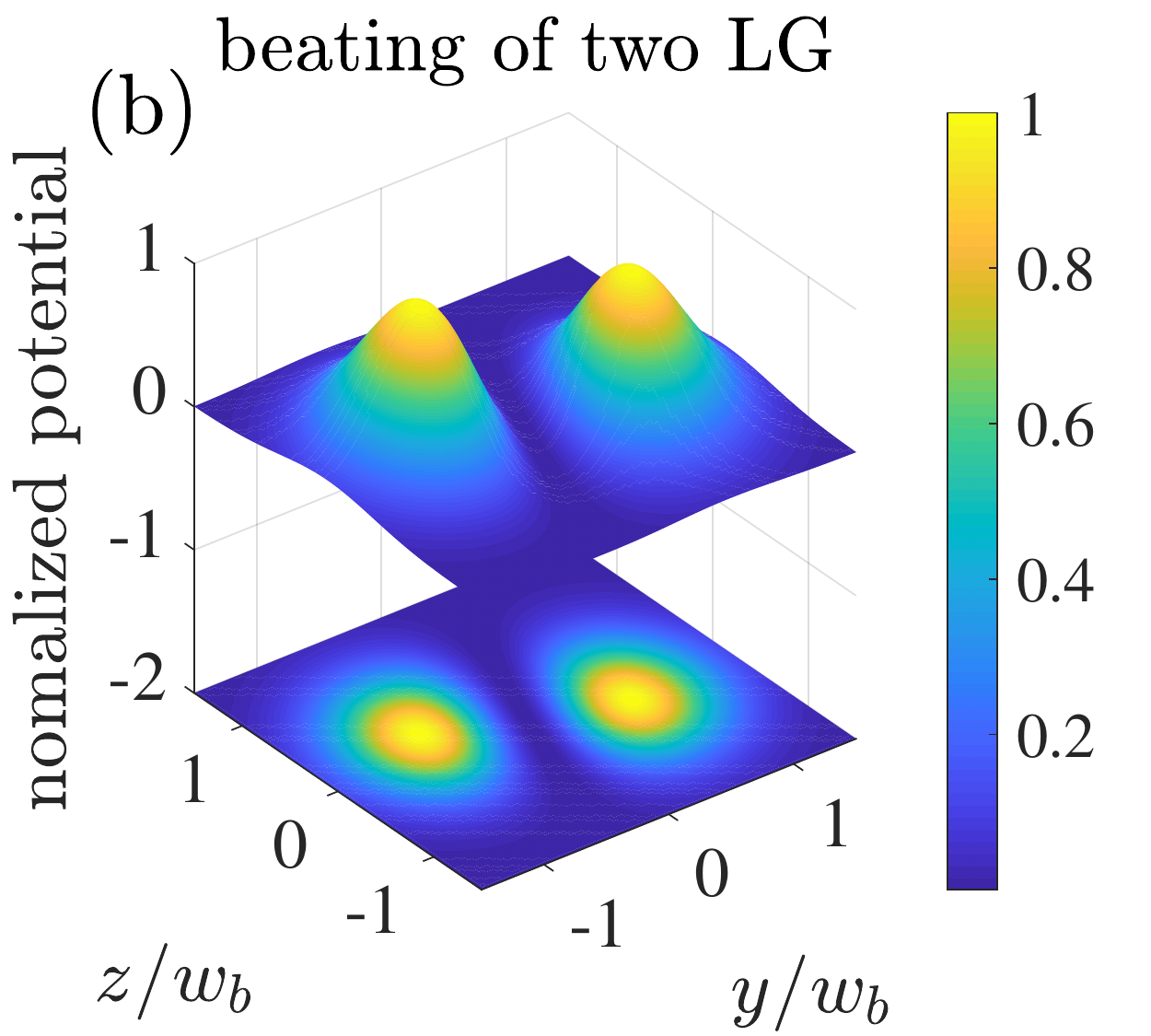}
	\caption{\small Structure of the ponderomotive potential $\Phi_{pond}$ (in a. u.) of one LG laser pulse (a) and two beating LG laser pulses (b) in transverse plane ($y$-$z$). The ponderomotive force ${\bf F}_{pond} = - \nabla \Phi_{pond}$ has an azimuthal component only for two beating waves. }
	\label{fig:twistforce}
\end{figure*}

In the context of the discussions about the OAM of light, the multiplier term of $\exp \left( -i l \theta \right)$ in Eq.~(\ref{e_lg}) is the most significant one. It sets the twist of the wavefronts and determines the OAM of an mode. The OAM per photon for this mode is $l \hbar$, where $l $ is the twist index. The laser pulse can efficiently exchange OAM with a structured target upon reflection~\cite{shi2014}, but the same cannot be said about a helical pulse propagating through a uniform plasma. This can be illustrated by examining the direction of the ponderomotive force induced by a helical beam with electric field amplitude ${\bf E}$ and frequency $\omega$ on plasma electrons,
\begin{eqnarray} \label{pond_pot}
    {\bf F}_{pond} = - \nabla \varphi_{pond}, &\mbox{    }&  \varphi_{pond} = \frac{e^2}{4 m_e \omega^2}  \left| {\bf E} ( {\bf r}, t) \right|^2,
\end{eqnarray}
where $\varphi_{pond}$ is the ponderomotive potential and $e$ and $m_e$ are the electron charge and mass. The intensity of $|E|^2$ is independent of $\theta$ and it follows from Eq.~(\ref{pond_pot}) that $\varphi_{pond}$ has no dependence on $\theta$ as well (see Fig.~\ref{fig:twistforce}(a)). Therefore, the ponderomotive force ${\bf F}_{pond}$ of a single LG beam has only a radial component in the cross section of the beam and it induces no azimuthal rotation of the plasma. The ponderomotive force is nonlinear, so the discussed difficulty can be overcome by using two co-propagating LG modes with different frequencies and twist indices to generate a ponderomotive force with a substantial azimuthal component.

We illustrate this for a pair of LG waves that have $p =0 $ and opposite twist indices, $l_1 = - l_2 = l$, so that they have the same radial dependence. The waves are beating with slightly different frequencies: $\omega_1 = \omega_0 $ and $\omega_2 = \omega_0 - \Delta \omega $, where $\Delta \omega \ll \omega_0$. 
 We keep the radial normalisation equivalent for all modes such that the real part of the final electric field looks like:
\begin{equation}
  \mathbf{E} = \mathrm{\hat{e_s}}E_0F_{p,l}(X)\left[\cos\left(k_1x -\omega_1 t -l\theta\right)+\cos\left(k_2x -\omega_2 t + l \theta\right)\right]
\end{equation}
Considering an average over the timescale $t_{fast}=2\pi/(\omega_1+\omega_2)$ and lengthscale $x_{fast}=2\pi/(k_1+k_2)$ we have a slowly varying electric field of the form:
\begin{equation}
  \mathbf{E}_{slow} = \mathrm{\hat{e_s}}2E_0F_{p,l}(X)\cos\left(-\frac{\Delta k x}{2} +\frac{\Delta\omega t}{2} + l\theta\right)
\end{equation}
where $\Delta k=k_1-k_2$. This slowly varying envelope of the total electric field retains all of the angular momentum of the two combined waves, it is worth mentioning that the wave travels in the same direction as the sum of its parts, despite the sign change in $k$.
To find the ponderomotive force we use the relation:
\begin{eqnarray}\label{Force_pond}
  \mathbf{F_{pond}} &=& -\frac{e^2}{4 m_e \omega^2}\nabla(E^2) \\
      &=& -\frac{e^2E^2_0}{2m_e \omega_0^2}\nabla\left[F^2_{p,l}(X)(1+\cos\left(-\Delta k x +\Delta\omega t +2 l\theta\right))\right].\nonumber
\end{eqnarray}
Putting $a_0=eE_0/(m_e \omega_0 c) $, the ponderomotive potential can be described as \begin{equation}
  \Phi_{pond}= 0.5m_e c^2 a_0^2F^2_{p,l}(X)(1+\cos\left(-\Delta k x +\Delta\omega t +2 l\theta\right))
\end{equation}
The ponderomotive potential is then no longer cylindrically symmetric (see Fig.~\ref{fig:twistforce}(b)).
The resulting ponderomotive force, ${\bf F}_{pond} = - \nabla \Phi_{pond}$, has an azimuthal component and it is slowly rotating in the cross section of the propagating beams, i.e. in the $y$-$z$ plane. In contrast to this, ${\bf F}_{pond}$ produced by a superposition of conventional beams without OAM ($l = 0$) has no azimuthal component. The remainder of this work will elucidate twisted plasma waves driven by the twisted ponderomotive force and axial magnetic field generation related to the axial OAM carried by electrons.

\section{\sffamily{Simulation results driven by twisted ponderomotive force}} \label{sec:SimRst}

Using EPOCH~\cite{Arber2015},  \rc{we} performed three-dimensional PIC simulations on the interactions between two beating LG-CP beams and plasma. For two LG-CP beams ($p = 0$), we choose the $l + \sigma =0$ to get zero angular momentum. So, the magnetic field generation due to the IF effect is excluded. 
Some key simulation parameters are summarized in Table \ref{table:PIC}. The two LG-CP beams have frequencies and twist indices of $\omega_1 = \omega_0$, $\omega_2 = 0.9\omega_0$ and $l_1$ = -1 , $l_2$ = +1, respectively. The frequency difference is set as the same as the plasma frequency $\omega_p=0.1\omega_0$.  $\omega_p^2 = 4 \pi n_0 e^2/m_e$ is the plasma frequency, where $n_0$ is the electron density and $\omega_0$ is the frequency of the beam 1 with 0.8~$\mu$m wavelength. The target is set as a fully ionized hydrogen plasma with uniform density $n_0$ = $4.5 \times 10^{18}$~cm$^{-3}$ and zero temperature, initially. Both the incident pulses have a Gaussian shape with a total duration of $\tau_g$ =200~fs with a beam width of $w_b = 10~\mu$m, and the electric field amplitude of $E_{y,z}$ = 1.0~TV/m. For the beam 1, the corresponding peak dimensionless vector potential of $a_0$ =0.25.

\begin{figure*}
\centering
\includegraphics[width=0.8\linewidth]{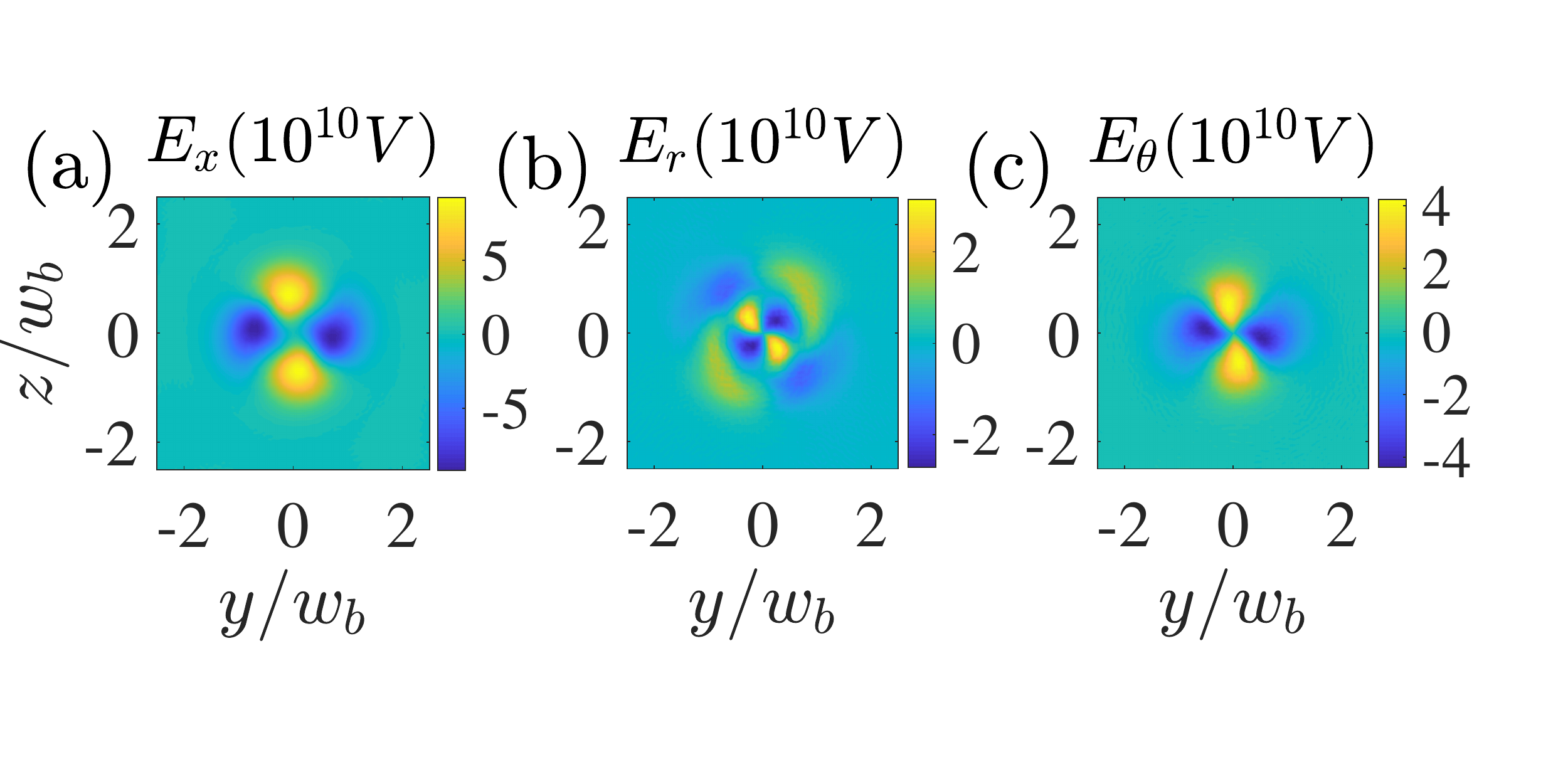} 
\caption{3D PIC simulation results of electric field and fluid velocity distribution at transverse plane ($y$-$z$ plane) at the centre of simulation box ($x$ = 15 $\mu$m) and the time 320~fs after the laser has passed by. (a), (b) and (c) show  transverse slices of $E_x$, $E_{\theta}$ and $E_r$.} \label{ev_sim}
\end{figure*}

\begin{figure*}
\centering
\includegraphics[width=0.9\columnwidth]{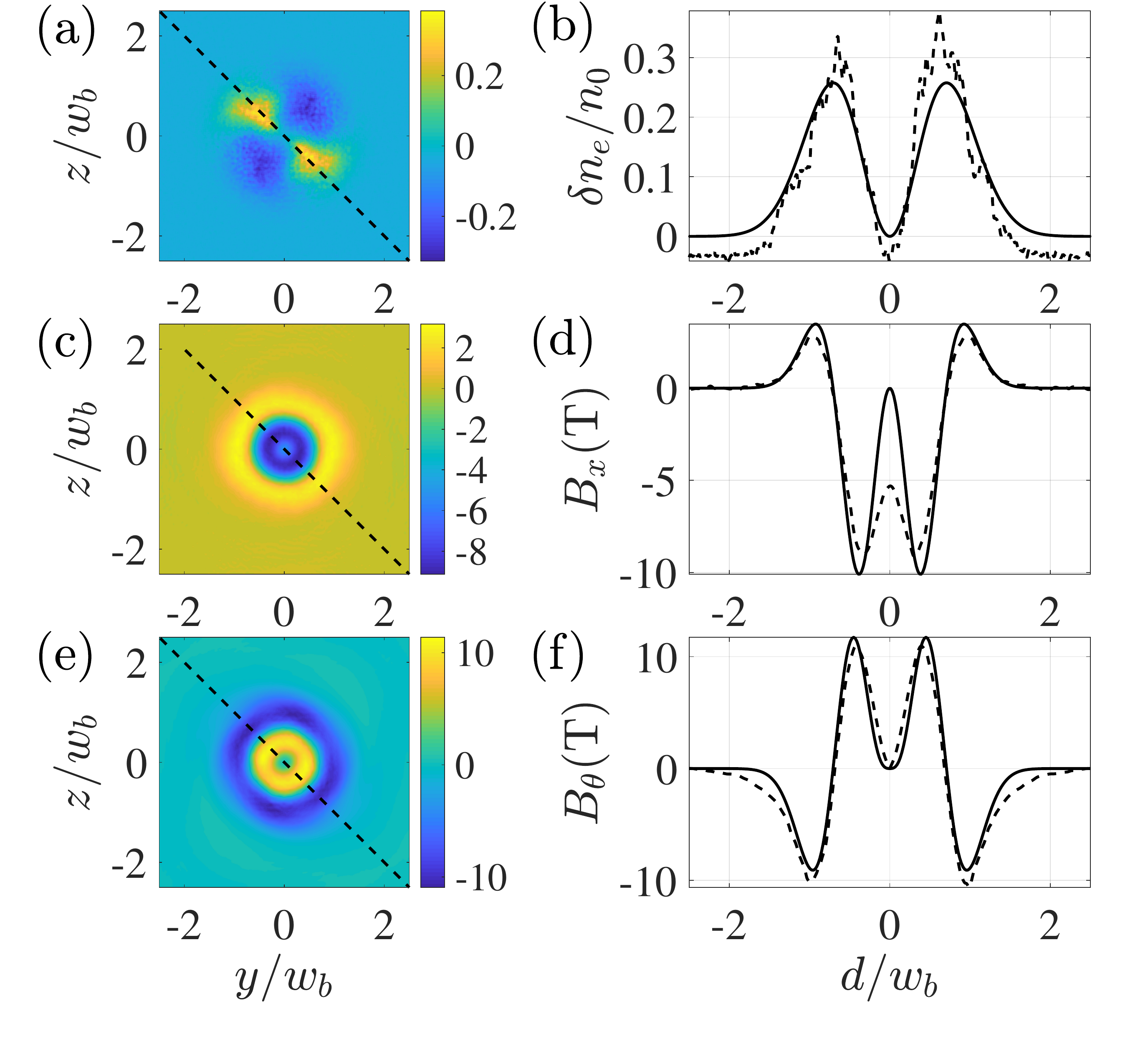}
\caption{PIC results of transverse profile of (a) electron density perturbation $\delta n_e$, (c) axial magnetic field $B_x$ and (e) azimuthal magnetic field $B_{\theta}$ at  the  centre  of  simulation  box ($x$ = 15 $\mu$m) and the time 320~fs after the laser has passed by.  The dashed lines shown in the transverse slices are the line outs used to plot the graphics on the right. The plots on the right, (b), (d), and (f), are line outs from the slices plotted against the position along the line outs $d$ plotted in (a), (c), and (e). $d$ is the coordinate along the dashed lines. Solid lines in (b), (d) and (f) are theory  predictions, for the same situation considered in table.~\ref{table:PIC}} \label{njb_sim}
\end{figure*}

In the followings, we will give the main simulation results. The twisted electric field ($E_r$, $E_{\theta}$, $E_x$) distribution are shown in Fig.~\ref{ev_sim}. We should notice that the non-zero distribution of $E_{\theta}$ (see Fig.~\ref{ev_sim}(b)) clearly shows characters of twist, which do not exist in normal longitudinal plasma waves. Furthermore, we find the phase distribution of  $E_r$ (see  Fig.~\ref{ev_sim}(a)),  and $E_x$ (see  Fig.~\ref{ev_sim}(c)) are twisted with azimuth $\theta$. These are different compared to longitudinal plasma waves driven by the beating of two Gaussian beams where there would be no dependence with $\theta$~\cite{Fedele1986}. Of particular note are the twisted electron density distribution $\delta n_e$ presented in Fig.~\ref{njb_sim}(a) and a static axial magnetic field $B_x$ in $y$-$z$ plane, shown in Fig.~\ref{njb_sim}(c). The magnetic field can be as high as 8T and persists for hundreds of femtoseconds after the laser beams pass by. We also observe an azimuthal magnetic field distribution which is presented in Fig.~\ref{njb_sim}(e). The dashed lines on the right of Fig.~\ref{njb_sim} are line-outs from the slices plotted against the position along the line-outs $d$ plotted in the left of Fig.~\ref{njb_sim}. As time moves on, the twisted electron density rotates around the $x$-axis with angular frequency $\omega_{pe}$. The scheme works for other choices of different twist indices (see 3D PIC simulation results in Section~\ref{sec:LgG}). There, twist effects using a LG and a Gaussian beam are confirmed and likely to be easier to realize experimentally. But to simplify the theoretical analysis, we choose opposite twist indices in the following analysis. We also choose circularly polarized lasers instead of linear polarized lasers to exclude the effects of polarization deflection due to the twisted electron density. The dependence of polarization deflection on the laser frequency can lead to a symmetry break in the transverse plane.  
\rc{Frequency detuning effects should be discussed for a real experiment. Additional simulations have been accomplished with different beating frequencies. We found that the axial magnetic fields are in similar distributions though a weaker peak amplitude ($\approx 90\%$) when the beating frequency is $\Delta \omega =  (1.00 \pm 0.05)\omega_p$.}

\begin{table}
\centering
\begin{tabular}{ |l|l|l| }
  \hline
\textbf{Laser parameters}  & \textbf{LG-CP beam1} & \textbf{LG-CP beam2} \\
  \hline
  Electric field amplitude($E_{pl}$) & $E_{1(y,z)}$= 1.0 TV/m & $E_{2(y,z)}$ = 1.0 TV/m \\
  Wavelength & $\lambda_1 = 0.8$ $\mu$m  & $\lambda_2 = 0.89$ $\mu$m\\
  twist index & $l_1$ = -1 & $l_2$ = 1\\
  Pulse duration (Gaussian shape) & $\tau_g = 200$ fs(75 cycles) & $\tau_g = 200$ fs(67.4 cycles) \\   Focal spot size ($1/\mathrm{e}$ electric field) & $w_b$ = 10 $\mu$m & $w_b$ = 10 $\mu$m\\
  Laser propagation direction & $+x$ & $+x$\\
  \hline \hline
  \multicolumn{3}{|l|}{\textbf{Other parameters} }\\
  \hline
  Electron density &\multicolumn{2}{|l|}{ $n_e = 0.01\;n_{c}$} \\  
 Simulation box ($x\times y \times z$) & \multicolumn{2}{|l|}{30 $\mu$m  $\times 50 $ $\mu$m $\times 50$ $\mu$m}\\  
  Cell number ($x\times y \times z$) & \multicolumn{2}{|l|}{600 $\times$ 800 $\times$ 800} \\
  Macroparticles per cell for each species & \multicolumn{2}{|l|}{2} \\
   \hline
   \end{tabular}
  \caption{3D PIC simulation parameters. $n_{c} = 1.8\times 10^{21}$~cm$^{-3}$ is the critical density corresponding to the laser wavelength 0.8 $\mu$m.}
  \label{table:PIC}
\end{table}

\section{\sffamily{The general theory of twisted plasma waves driven by twisted ponderomotive force}} \label{sec:theory}

 Firstly, we solve the non-relativistic cold electron-fluid equations driven by the twisted ponderomotive force in Section~\ref{sec:PondPotnl}. Using Linear Fluid Theory (LFT)~\cite{Fedele1986}, we can describe laser-driven electron plasma waves as 
\begin{equation}\label{LFT}
\left\{
             \begin{array}{lr}
                m_e \partial_t \mathbf{u} = e\nabla\Phi +\mathbf{F}_{pond} \\
                \partial_t \delta n_e +n_0 \nabla\cdot\mathbf{u} = 0 \\
                \nabla^2\Phi = 4 \pi e \delta n_e 
             \end{array}. 
\right.
\end{equation}
Here, ions are assumed to be immobile, $\delta n_e$ is the difference between the electron densities $n_e$  and ion density $n_0$. $\bf u$ is the velocity of the electron fluid and $\Phi$ is the static electric potential.  
The resulting equation for the perturbations of electron density $\delta n_e$ in a plasma with an initially uniform electron density $n_0$ has the form
\begin{equation}
  \partial_t^2 \delta n_e +\omega^2_{pe}\delta n_e = \frac{n_0}{m_e}\nabla\mathbf{F}_{pond}
\end{equation}
This equation is analogous to the equation that describes a driven harmonic oscillator. {The driving item can be calculated as 
\begin{equation}
\left\{
\begin{array}{lr}
  r^{-2}\partial^2_{\theta}\Phi_{pond} = -4a_0^2 l^2/(Xw_b^2) F^2_{p,l}(X)\cos\left(-\Delta k x +\Delta\omega t +2l\theta\right)\nonumber\\
   r^{-1}\partial_r(r\partial_r)\Phi_{pond}
  =4(a_0^2/w_b^2)\left[(F^2_{p,l}(X))' +X(F^2_{p,l}(X))'' \right][1+\cos\left(-\Delta k x +\Delta\omega t +2l\theta\right)]\nonumber\\
  \partial^2_x\Phi_{pond} =  -0.5\Delta k^2 a_0^2 F^2_{p,l}(X)\cos\left(-\Delta k x +\Delta\omega t +2l\theta\right)\nonumber
  \end{array}. 
\right.
\end{equation} }
With a definition of $G_{p,l}(X)$, 
\begin{equation}
  G_{p,l}(X) = \frac{(F^2_{p,l}(X))' +X(F^2_{p,l}(X))''}{F^2_{p,l}(X)}, 
\end{equation}
we can write $\nabla^2\Phi_{pond}$ as
\begin{eqnarray}\label{pond_full}
  \nabla^2\Phi_{pond} &=& \frac{a_0^2}{2} F^2_{p,l}(X)\left[-\Delta k^2-\frac{8 l^2}{Xw_b^2} + \frac{8}{w_b^2}G_{p,l}(X)\right]\cos\left(-\Delta k x +\Delta\omega t +2l\theta\right) \nonumber\\ &+&\frac{4a_0^2}{w_b^2} F^2_{p,l}(X)G_{p,l}(X).
\end{eqnarray}
We will only consider the part with the cosine which will drive our forced plasma oscillation. When the ponderomotive force drives the perturbations with a frequency equal to $\Delta \omega = \omega_{pe}$,we have a solution of $\delta n_e$ as 
\begin{eqnarray}\label{eDenst}
\frac{\delta n_e}{n_0} &=& \frac{\tau c^2\Delta k^2a_0^2}{4\omega_p} \Xi_{p,l}(X)\sin\left(-\Delta k x +\omega_{pe} t +2 l\theta\right), \\ \nonumber
\Xi_{p,l}(X) &=& F^2_{p,l}(X)\left[1+\frac{1}{\Delta k^2w_b^2}\frac{8 l^2}{X} - \frac{8}{\Delta k^2w_b^2}G_{p,l}(X)\right].
\end{eqnarray}
It suggests a linearly growing amplitude with the laser pulse duration $\tau$. Our simulation use two LG-LP beams having same parameters except different wavelengths and twist indexes ($p = 0, l_1 = - l_2 = l$).The function of $\Xi_{p,l}(X)$ electron density perturbation can be presented in the following explicit form,
 \begin{equation}
\Xi_{p,l}(X) =  F^2_{0,l}(X)\left[1 + \frac{8}{\Delta k^2w_b^2}( - X + 1 + 2|l|)\right]
\end{equation}
It is the same result with the Eq.(2) in paper~\cite{Shi2018}.  If $1/\Delta k^2 w_b^2$ is small then $\Xi_{p,l}(X) = F^2_{p,l}(X)$ for simple cases, and for cases where $p=0$ then $\Xi_{0,l}(X) = C_{0l}X^{|l|}e^{-X}/|l|!$, this can calculate a situation for a gaussian wave where $l=0$ so that $\Xi_{0,0}(X)=e^{-X}$. In this paper, we will focus on the case of paraxial approximation with small $1/\Delta k^2 w_b^2$.
 
In a similar way, the electric field $\mathbf{E}$ and the velocity field $\mathbf{u}$ can be solved in following equations,
\begin{equation}
\left\{
\begin{array}{lr}
   \partial_t^2 \mathbf{E} +\omega_{pe}^2\mathbf{E} = -\omega_{pe}^2\mathbf{F_{pond}}/e \\ 
   \partial_t^2 \mathbf{u} +\omega_{pe}^2\mathbf{u} = \partial_t\mathbf{F_{pond}}/m_e
  \end{array} , 
\right.
\end{equation}
According to the Eq.~\ref{Force_pond}, the force is
\begin{eqnarray}
\left\{
\begin{array}{lr}
  F_{pond,\theta} = (l c^2 a_0^2m_e/w_b)\sqrt{2/X} F^2_{p,l}(X)\sin\left(-\Delta k x +\Delta\omega t +2 l\theta\right)\\
  F_{pond,r} = -(c^2 a_0^2m_e/w_b)\sqrt{2X}(F^2_{p,l}(X))'\cos\left(-\Delta k x +\Delta\omega t +2 l\theta\right) \\
    F_{pond,x} = -0.5(c^2 a_0^2m_e\Delta k) F^2_{p,l}(X)\sin\left(-\Delta k x +\Delta\omega t +2 l\theta\right)
   \end{array}.
\right.
\end{eqnarray}
We can ignore the last part of the constant force in the radial direction. For the velocity we can again neglect the constant radial term giving $\partial_t F$ as,
\begin{eqnarray}
\left\{
\begin{array}{lr}
  \partial_tF_{pond,\theta} = (l c^2 a_0^2m_e\omega_{pe}/w_b)\sqrt{2/X} F^2_{p,l}(X)\cos\left(-\Delta k x +\Delta\omega t +2 l\theta\right)\\
  \partial_tF_{pond,r} = (c^2a_0^2m_e\omega_{pe}/w_b)\sqrt{2X}(F^2_{p,l}(X))'\sin\left(-\Delta k x +\Delta\omega t +2 l\theta\right)\\
 \partial_t F_{pond,x} = -0.5(c^2a_0^2m_e\omega_{pe}\Delta k) F^2_{p,l}(X)\cos\left(-\Delta k x +\Delta\omega t +2 l\theta\right)
 \end{array}.
\right.
\end{eqnarray}
Mirroring the approach used above, the electric field and fluid velocity are found to be
 \begin{eqnarray}\label{E3d}
 \left\{
\begin{array}{lr}\label{Efield}
  E_{\theta} = -0.5\tau\omega_{pe} l c^2 a_0^2(m_e/e/w_b)\sqrt{2/X} F^2_{p,l}(X)\cos\left(-\Delta k x +\Delta\omega t +2 l\theta\right)\\
  E_{r} = -\tau\omega_{pe}(c^2 a_0^2m_e/e/w_b)\sqrt{X/2}(F^2_{p,l}(X))'\sin\left(-\Delta k x +\Delta\omega t +2 l\theta\right) \\
    E_{x} = 0.25\tau\omega_{pe}c^2 a_0^2(m_e/e)\Delta k F^2_{p,l}(X)\cos\left(-\Delta k x +\Delta\omega t +2 l\theta\right)
 \end{array} , 
\right.
\end{eqnarray}
and 
\begin{eqnarray}\label{u3d}
 \left\{
\begin{array}{lr}
  u_{\theta} =  -0.5(l c^2 a_0^2\tau/w_b) \sqrt{2/X} F^2_{p,l}(X)\sin\left(-\Delta k x +\Delta\omega t +2 l\theta\right)\\
 u_{r} = (c^2 a_0^2\tau/w_b)\sqrt{X/2}(F^2_{p,l}(X))'\cos\left(-\Delta k x +\Delta\omega t +2 l\theta\right)\\
   u_{x} = 0.25(c^2a_0^2\tau\Delta k) F^2_{p,l}(X)\sin\left(-\Delta k x +\Delta\omega t +2 l\theta\right)
 \end{array}. 
\right.
\end{eqnarray}
Using the explicit form of  $(F^2_{0, l}(X))' = (|l|/X - 1)F^2_{0, l}$, we can get the same result of Eq.(s7) and Eq.(s8) in the Supplemental Material of paper~\cite{Shi2018}.
 
In theory above,  the electric current density is ${\bf j}^e = -e n_0 {\bf u}$ , and the displacement current density is  ${\bf j}^{dis} = \partial {\bf E}/(4\pi \partial t)$. According to high order fluid theory~\cite{Gorbunov1996,Gorbunov1997}, the second-order current can generate magnetic field even though  ${\bf j}^e + {\bf j}^{dis} = 0$ in the first order. In PIC simulation results, time-averaged net current in both the azimuthal direction  $\langle j^{(2)}_{\theta} \rangle$ (i.e. net rotating current) and the axial direction $\langle j^{(2)}_{x} \rangle$ are confirmed. With a definition of $\widetilde{j_0}= -en_0\tau^2 c^4\Delta k^3a_0^4/\omega_p$, 
the second-order current is  
\begin{eqnarray}
 \left\{
\begin{array}{lr}
  j^{(2)}_{\theta} =  -(l/8) \widetilde{j_0}/(w_b\Delta k) \sqrt{2/X} F^4_{p,l}(X)\sin^2\left(-\Delta k x +\Delta\omega t +2 l\theta\right)\\
  j^{(2)}_{r} = (1/8)\widetilde{j_0}/(w_b\Delta k)\sqrt{X/2}(F^2_{p,l}(X))'F^2_{p,l}(X)\sin2\left(-\Delta k x +\Delta\omega t +2 l\theta\right)\\
    j^{(2)}_{x} = (1/16)\widetilde{j_0} F^4_{p,l}(X)\sin^2\left(-\Delta k x +\Delta\omega t +2 l\theta\right)
 \end{array}.
\right.
\end{eqnarray} 
The equation for the second-order vector potential $\mathbf{A}^{(2)}$ can be written as 
\begin{equation}
    (\partial_t^2 - c^2\nabla^2 + \omega_{pe}^2)\mathbf{A}^{(2)} = 4\pi c \mathbf{j}^{(2)}
\end{equation}
In the paraxial approximation where only the dominant axial derivative in the Laplacian term is accounted, we can get the explicit expressions for the vector potential  $\mathbf{A}^{(2)}$ as
 \begin{eqnarray}
 \left\{
\begin{array}{lr}
  A^{(2)}_{\theta} =  -(l\pi c/4) \widetilde{j_0}/(w_b\Delta k) \sqrt{2/X} F^4_{p,l}(X)[\frac{1}{\Delta\omega^2} -\frac{1}{4\Delta k^2c^2 - 3\Delta\omega^2}  \cos2\left(-\Delta k x +\Delta\omega t +2 l\theta\right)]\\
  A^{(2)}_{r} = (\pi c/2)\widetilde{j_0}/(w_b\Delta k)\sqrt{X/2}(F^2_{p,l}(X))'F^2_{p,l}(X)\frac{1}{4\Delta k^2c^2 - 3\Delta\omega^2}\sin2\left(-\Delta k x +\Delta\omega t +2 l\theta\right)\\
    A^{(2)}_{x} = (\pi c/8)\widetilde{j_0} F^4_{p,l}(X)[\frac{1}{\Delta\omega^2} -\frac{1}{4\Delta k^2c^2 - 3\Delta\omega^2}  \cos2\left(-\Delta k x +\Delta\omega t +2 l\theta\right)]
 \end{array}. 
\right.
\end{eqnarray} 
The axial and azimuthal components contains constant term and second harmonics oscillating term. But the radial component has only second harmonics oscillating term. The quasi-stationary magnetic field can be calculated by $\mathbf{B} = \nabla \times \mathbf{A}^{(2)}$, 
\begin{eqnarray}\label{B3d}
 \left\{
\begin{array}{lr}
B_{r} = \frac{1}{r}\frac{\partial A^{(2)}_x}{\partial \theta} - \frac{\partial A^{(2)}_{\theta}}{\partial x} = 0 \\
B_{\theta} =  - \frac{\partial A^{(2)}_{x}}{\partial r} = -\frac{\pi c\widetilde{j_0}}{4w_b\Delta\omega^2} \sqrt{X}(F^4_{p,l})'\\
B_{x} = \frac{1}{r}\frac{\partial rA^{(2)}_{\theta}}{\partial r} = -\frac{l \pi c\widetilde{j_0}}{w_b^2\Delta\omega^2\Delta k} (F^4_{p,l})'
\end{array}.
\right.
\end{eqnarray} 
The relationship between the azimuthal and axial component is $B_{\theta}/B_{x} = 0.25(w_b\Delta k/l)\sqrt{X}$.
 With the parameters of $\kappa_w = \Delta k w_b$ and $\kappa_n = \tau c^2\Delta k^2a_0^2/(4\omega_p)$, we can get the amplitude of $B_x$, 
\begin{eqnarray}\label{Bx_scale}
   B_{x0} = -\frac{l \pi c\widetilde{j_0}}{w_b^2\Delta\omega^2\Delta k} =  8 l c\sqrt{\pi m_e n_0}(\frac{\kappa_n}{\kappa_w})^2.
\end{eqnarray}  
If we have $\kappa_w = 10$, $\kappa_n = 0.1$ and $\widetilde{n_0} = n_0 / n_c$, we can get $B_{x0}$(T) $ \sim 5 l \sqrt{\widetilde{n_0}}$. It means that if we assume that the electron density perturbation is small and the paraxial approximation works well, the amplitude of $B_x$ will increase with bigger plasma density. Since the plasma density can not be overdense in our scheme here, we expect there is an upper limit on the amplitude of $B_x$.

In the PIC simulation shown in Table~\ref{table:PIC}, Fig.~\ref{ev_sim} and Fig.~\ref{njb_sim} give the results. Here, $1/w_b^2\Delta k^2 = 0.016$ when we use $\Delta k = \omega_{pe}/c$. For simplicity, Eq.~(\ref{eDenst}) has been obtained assuming a flat-top pulse with duration $\tau$. For the truncated Gaussian pulse used in the PIC simulation (i.e. $I(t)$ = 0 for $|t - t_p| \ge \tau_g/2$ where $t_p= \tau_g/2$ is the time of peak intensity), $\tau$ =0.75$\tau_g$ is appropriate. The Eq.~(\ref{E3d}) show that the radial field $E_r$ is phase different compared to the azimuthal field $E_{\theta}$ and axial fields $E_{x}$. These are the same with results in Fig.~\ref{ev_sim}. The radial field $u_r$ is also phase different compared to the azimuthal field $u_{\theta}$ and axial fields $u_{x}$ according to Eq.~(\ref{u3d}). 
Both $u_{\theta} \propto l$ and  $E_{\theta} \propto l$ show the importance of the OAM of the driving lasers.  
The calculation results of electron density perturbation in the transverse plane are shown as a solid line in Fig.~\ref{njb_sim}(b), which is close to the simulation result (dashed line). The difference is thought to be caused by ignoring the last part of the constant force in the radial direction in Eq.~(\ref{pond_full}). Theory predictions of magnetic field from Eq.~(\ref{B3d}) are shown as solid lines in Fig.~\ref{njb_sim}(d) and (f). They are also close to the dashed lines. The result of the LFT is twisted plasma waves with a helical rotating structure. It is different from the longitudinal plasma waves driven by the beating of two Gaussian lasers, where transverse profiles depend only on radius~\cite{Fedele1986}. The magnetic field generation due to the second-order current is explained well in the paraxial approximation. {While this study is limited to plasma waves with small amplitude, the high order terms are ignored in Eq.~(\ref{LFT}). In other simulations, we observed spiral phenomena which may be related to the high-order nonlinear terms.}

To understand more details on the twisted plasma waves with electrons carrying axial OAM, we analysis the oscillating phase of single electron under a transverse electric field of $E_r = E_{r1}\sin \Phi_0$ and $E_{\theta} = E_{\theta 1} \cos \Phi_0$, where $\Phi_0 = \omega_p t + 2l\theta_0$ at $x = 0$ and $\theta_0$ is the azimuth.  Here, $E_r(r_0, \theta_0)$ and $E_{\theta}(r_0, \theta_0)$ are rewrote for our purpose according to Eq.~\ref{Efield}.  The solutions of  the motion equation  $m_e d^2 {\bf r}_1/dt^2 = -e{\bf E}(t, {\bf r}_0)$ to the first order in coordinate of $(y, z)$  are 
\begin{equation}\label{KntMotion}
\left\{
             \begin{array}{lr}
            v_{y1} = (E_{r1}\cos\theta_0 \cos\Phi_0 + E_{\theta 1}\sin\theta_0 \sin\Phi_0 )(e/m_e\omega_p)  &  \\
            v_{z1} = (E_{r1}\sin\theta_0 \cos\Phi_0 - E_{\theta 1}\cos\theta_0 \sin\Phi_0 )(e/m_e\omega_p)  &  
             \end{array}. 
\right.
\end{equation}
 Within such an oscillating electric field, electrons will oscillate locally with an azimuthally dependent phase to the first order. The central shift velocity is assumed small and should have no contribution to the rotating phase $\phi$. If the amplitude dependence on radius is  kept only for positive and negative, the dependence of rotating phase $\phi$ on azimuth $\theta_0$ is 
\begin{equation}\label{PhsMotion}
\phi = \arctan(\frac{v_{z1}}{v_{y1}})=
\left\{
             \begin{array}{lr}
           \omega_p t + (2l +1)\theta_0, & r_0 < r_b  \\
        -\omega_p t - (2l -1)\theta_0, & r_0 > r_b  
             \end{array},
\right.
\end{equation}
where $r_b = w_b\sqrt(|l|/2)$ and $l$ = -1. Furthermore, we can show that the rotating electron density in transverse plane comes from particles oscillating elliptically in the transverse plane with an azimuthally dependent phase. We reconstruct the 2D electron density  perturbation in the transverse plane to a higher order approximation from particles oscillation. This oscillation can be viewed as a mapping of the $\bm{r}_0$ space onto the $\bm{r}_0 + \bm{r}_1$ space. Here, we have $\bm{r}_1 = \int (v_{y1}, v_{z1}) dt$. The mapping is assumed to be single-valued and regular here. Then, the number density due to the motion can be given by $ndV = n_0dV_0$, where the two volume elements ($dV_0$ and $dV$) are related by the Jacobian of the transformation ~\cite{Dawson1959}. From the electron oscillation, we can get 
\begin{equation}\label{Dawson}
   \frac{n_0}{n} = \frac{\partial(y_0 + y_1, z_0 + z_1)}{\partial(y_0, z_0)}.
\end{equation}
The electron density can be approximated to second order
\begin{eqnarray} \label{eDenst1}
\begin{aligned}
    n \approx n_0[1 +& \tilde{n}_1(\eta, r_0) \sin \Phi_0 + o(\eta^2, r_0) + o(\eta^2, r_0)\cos2\Phi_0],\\ & \tilde{n}_1 = \frac{1}{\omega_p}(2 l \frac{v_{\theta_1}}{r_0} + \frac{v_{r_1}}{r_0} +\frac{dv_{r_1}}{dr_0}).
\end{aligned}
\end{eqnarray}
Here, we get $v_{r1} = eE_{r1}/(m_e\omega_p)$ and  $v_{\theta1} = -eE_{\theta1}/(m_e\omega_p)$ from Eq.~(\ref{KntMotion}).  The explicit form of Eq.~\ref{eDenst1} shows that the first order electron density is the same as LFT result of Eq.~\ref{eDenst} in transverse plan if we ignore $k_p x$.  We can get the electric field in azimuth component to a second order, $E_{\theta} = E_{\theta 1}\cos \Phi_0 + o(\eta^2, r_0)\sin 2\Phi_0$. To the first order, it is just the same as the electric field from LFT. We can write a demo with some randomly distributed particles rotating with a phase dependence of azimuth in the plane. The collected motion of rotating with a phase dependence of azimuth will create a rotating electron density and carry axial OAM. Such a rotating electron density in plasma will produce a twisted electrostatic field, which in reality can push every electron rotating with a phase dependence of azimuth. It is a self-consistency process in twisted plasma waves.

At last, using the equation of $\langle L_x^e \rangle = -\langle j^{(2)}_{\theta} \rangle r m_e /e$, we can get the theoretical result of the transverse profile of the time-averaged axial OAM density.  The integration of $\langle L_x^e \rangle$ over radius $r$ is not zero when $l \ne 0$. For electromagnetic fields, the energy and momentum of electromagnetic waves are transported essentially by electric and magnetic fields. But for the plasma wave fields,  electrostatic waves in plasma carry energy but no momentum if we do not consider the magnetic field due to the high order effect. Only particles will carry axial momentum and OAM. The same results can also be found in paper~\cite{Blackman2019b}. In paper~\cite{Blackman2019a, Blackman2019b}, Langmuir plasma waves carrying a finite OAM are studied using kinetic theory in the paraxial approximation. Dispersion relation and energy, momentum exchange between particles and field in damping process are studied for the twisted plasma waves.

\section{\sffamily{Simulation results driven by beating between a LG pulse and a Gaussian pulse}} \label{sec:LgG}

\begin{figure*}
\centering
\includegraphics[width=0.9\columnwidth]{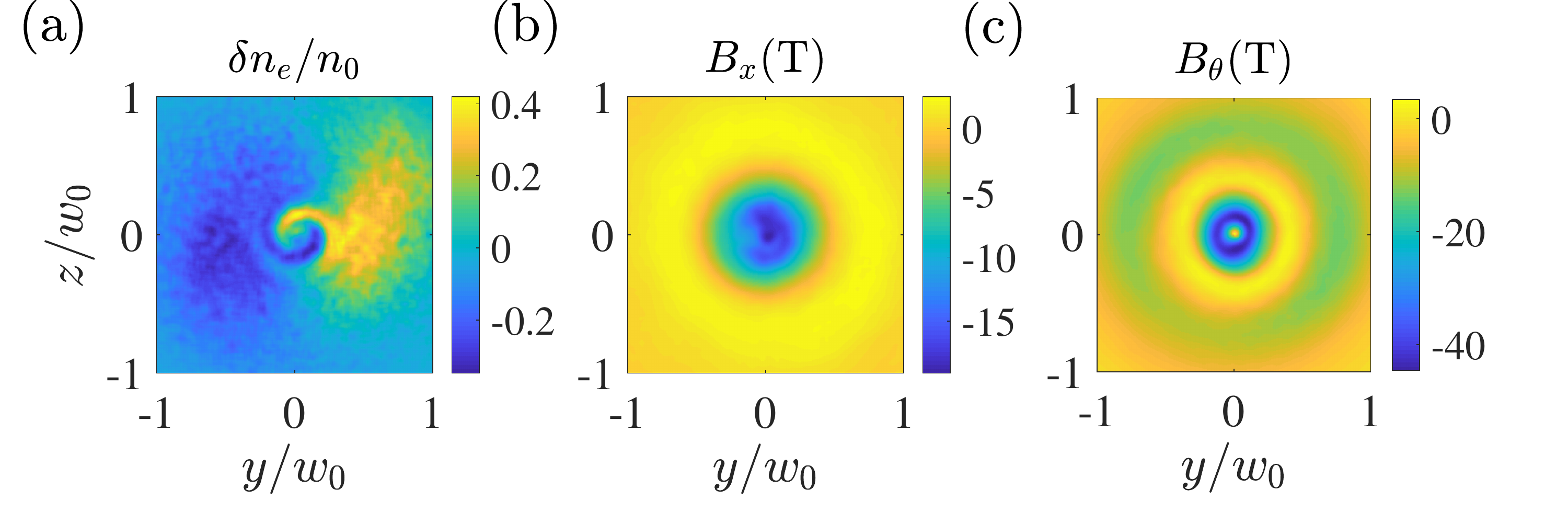}
\caption{PIC results of transverse profile of (a) electron density perturbation $\delta n_e$, (b) axial magnetic field $B_x$ and (c) azimuthal magnetic field $B_{\theta}$ at  the  centre  of  simulation  box ($x$ = 15 $\mu$m) and the time 320~fs after the laser has passed by.  } \label{TwistB_glg}
\end{figure*}

Considering the difficulties of producing high intense LG beams in experiment, we accomplished a simulation using the beating of one LG beam ($l$ = -1) and one Gaussian beam ($l$ = 0). Now, the ponderomotive force is ${\bf F}_{pond}(r, \theta, x, t) \propto \nabla \cos(\omega_p t - k_px - \theta)$ and then the electron density perturbation is $\delta n_e \propto \sin(\omega_p t - k_px - \theta)$. We set the beam 2 in Table~\ref{table:PIC} as a Gaussian beam and other parameters are kept the same.  Their  frequency  shift  was  also set  to  match  the  plasma frequency. The similar twisted electron density and magnetic field generation are observed in Fig.~\ref{TwistB_glg}. Of particular note are the single helical electron density distribution $\delta n_e$ presented in Fig.~\ref{TwistB_glg}(a). In Fig.~\ref{TwistB_glg}(b), the magnetic field can be as high as 20T. {In future experiments, we may consider using temporally-chirped pulses for twisted plasma waves excitation ~\cite{Cowley2017}}

\section{\sffamily{Summary}} \label{sec:summary}

In conclusion, twisted plasma waves carrying OAM are produced in PIC simulations by using beating of two LG beams or one LG and one Gaussian beam.  We compare the ponderomotive force of two beating LG beams with the ponderomotive force of two beating Gaussian beams. The substantial azimuthal component of ponderomotive force provides a unique tool for an effective exchange of OAM between the light and the plasma. Twisted electron plasma waves with a rotating helical structure are confirmed using Three-dimensional PIC simulations. Once being excited, the twisted plasma wave can persist in the plasma long after the laser beams that created the ponderomotive force are gone. Essentially, the electrons are left oscillating along elliptical orbits in the transverse plane with an azimuthally dependent phase offset. This collectively yields a persistent, rotating wave structure and an associated with it nonlinear electron current. The current configuration that is similar to that in a solenoid creates a longitudinal quasi-static magnetic field. Under the condition of paraxial approximation, we develop a general theory to explain the twist of the plasma waves and calculate the distributions of the magnetic fields in axial and azimuthal directions. At last, to relax the experimental conditions of getting relativistic twisted laser intensities, we run simulations using beating between a LG and a Gaussian beam. The twisted ponderomotive force and twisted plasma waves would be easier to realize experimentally when only one beam with OAM is required.


\section*{Data availability}

    The datasets generated during and analyzed during the current study are available from the corresponding author on reasonable request.
    
\section*{Code availability}

    PIC simulations were performed with the fully relativistic open-access 3D PIC code EPOCH\cite{EpochGit}.


\bibliographystyle{ieeetr}

\begin{thebibliography}{10}

\bibitem{Esarey2009}
E.~Esarey, C.~B. Schroeder, and W.~P. Leemans, ``Physics of laser-driven
  plasma-based electron accelerators,'' {\em Reviews of Modern Physics},
  vol.~81, pp.~1229--1285, 2009.

\bibitem{Ali2010}
S.~Ali, J.~R. Davies, and J.~T. Mendonca, ``Inverse faraday effect with
  linearly polarized laser pulses,'' {\em Physical Review Letters}, vol.~105,
  no.~3, p.~035001, 2010.

\bibitem{Haines2001}
M.~G. Haines, ``Generation of an axial magnetic field from photon spin,'' {\em
  Physical Review Letters}, vol.~87, no.~13, p.~135005, 2001.

\bibitem{Najmudin2001}
Z.~Najmudin, M.~Tatarakis, A.~Pukhov, E.~L. Clark, R.~J. Clarke, A.~E. Dangor,
  J.~Faure, V.~Malka, D.~Neely, M.~I.~K. Santala, and K.~Krushelnick,
  ``Measurements of the inverse faraday effect from relativistic laser
  interactions with an underdense plasma,'' {\em Physical Review Letters},
  vol.~87, no.~21, p.~215004, 2001.

\bibitem{Sheng1996}
Z.~M. Sheng and J.~Meyer-ter Vehn, ``Inverse faraday effect and propagation of
  circularly polarized intense laser beams in plasmas,'' {\em Physical Review
  E}, vol.~54, no.~2, pp.~1833--1842, 1996.

\bibitem{Allen1992}
L.~Allen, M.~W. Beijersbergen, R.~J.~C. Spreeuw, and J.~P. Woerdman, ``Orbital
  angular momentum of light and the transformation of laguerre-gaussian laser
  modes,'' {\em Physical Review A}, vol.~45, no.~11, pp.~8185--8189, 1992.

\bibitem{Yao2011}
A.~M. Yao and M.~J. Padgett, ``Orbital angular momentum: origins, behavior and
  applications,'' {\em Advances in Optics and Photonics}, vol.~3, no.~2,
  pp.~161--204, 2011.

\bibitem{shi2014}
Y.~Shi, B.~Shen, L.~Zhang, X.~Zhang, W.~Wang, and Z.~Xu, ``Light fan driven by
  a relativistic laser pulse,'' {\em Physical Review Letters}, vol.~112,
  no.~23, p.~235001, 2014.

\bibitem{vieira2016}
J.~Vieira, R.~M.~M. Trines, E.~P. Alves, R.~A. Fonseca, J.~T. Mendonça,
  R.~Bingham, P.~Norreys, and L.~O. Silva, ``High orbital angular momentum
  harmonic generation,'' {\em Physical Review Letters}, vol.~117, no.~26,
  p.~265001, 2016.

\bibitem{zhang2016}
L.~Zhang, B.~Shen, X.~Zhang, S.~Huang, Y.~Shi, C.~Liu, W.~Wang, J.~Xu, Z.~Pei,
  and Z.~Xu, ``Deflection of a reflected intense vortex laser beam,'' {\em
  Physical Review Letters}, vol.~117, no.~11, p.~113904, 2016.

\bibitem{Zhang2015}
X.~Zhang, B.~Shen, Y.~Shi, X.~Wang, L.~Zhang, W.~Wang, J.~Xu, L.~Yi, and Z.~Xu,
  ``Generation of intense high-order vortex harmonics,'' {\em Physical Review
  Letters}, vol.~114, no.~17, p.~173901, 2015.

\bibitem{vieira2014}
J.~Vieira and J.~T. Mendonça, ``Nonlinear laser driven donut wakefields for
  positron and electron acceleration,'' {\em Physical Review Letters},
  vol.~112, no.~21, p.~215001, 2014.

\bibitem{wang2015}
W.~Wang, B.~Shen, X.~Zhang, L.~Zhang, Y.~Shi, and Z.~Xu, ``Hollow screw-like
  drill in plasma using an intense laguerre–gaussian laser,'' {\em Science
  Report}, vol.~5, 2015.

\bibitem{Zhang2014}
X.~Zhang, B.~Shen, L.~Zhang, J.~Xu, X.~Wang, W.~Wang, L.~Yi, and Y.~Shi,
  ``Proton acceleration in underdense plasma by ultraintense
  laguerre–gaussian laser pulse,'' {\em New Journal of Physics}, vol.~16,
  no.~12, p.~123051, 2014.

\bibitem{vieira2018}
J.~Vieira, J.~T. Mendon\ifmmode~\mbox{\c{c}}\else \c{c}\fi{}a, and
  F.~Qu\'er\'e, ``Optical control of the topology of laser-plasma
  accelerators,'' {\em Phys. Rev. Lett.}, vol.~121, p.~054801, Jul 2018.

\bibitem{Longman2017}
A.~Longman and R.~Fedosejevs, ``Mode conversion efficiency to laguerre-gaussian
  oam modes using spiral phase optics,'' {\em Opt. Express}, vol.~25,
  pp.~17382--17392, Jul 2017.

\bibitem{Ju2018}
L.~B. Ju, C.~T. Zhou, K.~Jiang, T.~W. Huang, H.~Zhang, T.~X. Cai, J.~M. Cao,
  B.~Qiao, and S.~C. Ruan, ``Manipulating the topological structure of
  ultrarelativistic electron beams using laguerre{\textendash}gaussian laser
  pulse,'' {\em New Journal of Physics}, vol.~20, p.~063004, jun 2018.

\bibitem{Zhu2019}
X.-L. Zhu, M.~Chen, S.-M. Weng, P.~McKenna, Z.-M. Sheng, and J.~Zhang,
  ``Single-cycle terawatt twisted-light pulses at midinfrared wavelengths above
  10 $\mu$m,'' {\em Phys. Rev. Applied}, vol.~12, p.~054024, Nov 2019.

\bibitem{TIKHONCHUK2020}
V.~Tikhonchuk, P.~Korneev, E.~Dmitriev, and R.~Nuter, ``Numerical study of
  momentum and energy transfer in the interaction of a laser pulse carrying
  orbital angular momentum with electrons,'' {\em High Energy Density Physics},
  vol.~37, p.~100863, 2020.

\bibitem{Nuter2018}
R.~Nuter, P.~Korneev, I.~Thiele, and V.~Tikhonchuk, ``Plasma solenoid driven by
  a laser beam carrying an orbital angular momentum,'' {\em Phys. Rev. E},
  vol.~98, p.~033211, Sep 2018.

\bibitem{Blackman2020}
D.~R. Blackman, R.~Nuter, P.~Korneev, and V.~T. Tikhonchuk, ``Nonlinear landau
  damping of plasma waves with orbital angular momentum,'' {\em Phys. Rev. E},
  vol.~102, p.~033208, Sep 2020.

\bibitem{Longman2021}
A.~Longman and R.~Fedosejevs, ``Kilo-tesla axial magnetic field generation with
  high intensity spin and orbital angular momentum beams,'' {\em Phys. Rev.
  Research}, vol.~3, p.~043180, Dec 2021.

\bibitem{Leblanc2017}
A.~Leblanc, A.~Denoeud, L.~Chopineau, G.~Mennerat, P.~Martin, and F.~Quere,
  ``Plasma holograms for ultrahigh-intensity optics,'' {\em Nature Physics},
  vol.~advance online publication, 2017.

\bibitem{Denoeud2017}
A.~Denoeud, L.~Chopineau, A.~Leblanc, and F.~Quéré, ``Interaction of
  ultraintense laser vortices with plasma mirrors,'' {\em Physical Review
  Letters}, vol.~118, no.~3, p.~033902, 2017.

\bibitem{Longman2020}
A.~Longman, C.~Salgado, G.~Zeraouli, J.~I.~A. {n}aniz, J.~A.
  P\'{e}rez-Hern\'{a}ndez, M.~K. Eltahlawy, L.~Volpe, and R.~Fedosejevs,
  ``Off-axis spiral phase mirrors for generating high-intensity optical
  vortices,'' {\em Opt. Lett.}, vol.~45, pp.~2187--2190, Apr 2020.

\bibitem{BAE2020}
J.~Y. Bae, C.~Jeon, K.~H. Pae, C.~M. Kim, H.~S. Kim, I.~Han, W.-J. Yeo,
  B.~Jeong, M.~Jeon, D.-H. Lee, D.~U. Kim, S.~Hyun, H.~Hur, K.-S. Lee, G.~H.
  Kim, K.~S. Chang, I.~W. Choi, C.~H. Nam, and I.~J. Kim, ``Generation of
  low-order laguerre-gaussian beams using hybrid-machined reflective spiral
  phase plates for intense laser-plasma interactions,'' {\em Results in
  Physics}, vol.~19, p.~103499, 2020.

\bibitem{Aboushelbaya2020}
R.~Aboushelbaya, K.~Glize, A.~F. Savin, M.~Mayr, B.~Spiers, R.~Wang,
  N.~Bourgeois, C.~Spindloe, R.~Bingham, and P.~A. Norreys, ``Measuring the
  orbital angular momentum of high-power laser pulses,'' {\em Physics of
  Plasmas}, vol.~27, no.~5, p.~053107, 2020.

\bibitem{Xu2020}
X.~Zeng, S.~Zheng, Y.~Cai, H.~Wang, X.~Lu, H.~Wang, J.~Li, W.~Xie, and S.~Xu,
  ``Generation and imaging of a tunable ultrafast intensity-rotating optical
  field with a cycle down to femtosecond region,'' {\em High Power Laser
  Science and Engineering}, vol.~8, p.~e3, 2020.

\bibitem{Shi2018}
Y.~Shi, J.~Vieira, R.~M. G.~M. Trines, R.~Bingham, B.~F. Shen, and R.~J.
  Kingham, ``Magnetic field generation in plasma waves driven by copropagating
  intense twisted lasers,'' {\em Physical Review letters}, vol.~121, p.~145002,
  Oct 2018.

\bibitem{Blackman2019a}
D.~R. Blackman, R.~Nuter, P.~Korneev, and V.~T. Tikhonchuk, ``Kinetic plasma
  waves carrying orbital angular momentum,'' {\em Physical Review E}, vol.~100,
  p.~013204, Jul 2019.

\bibitem{Blackman2019b}
D.~R. Blackman, R.~Nuter, P.~Korneev, and V.~T. Tikhonchuk, ``Twisted kinetic
  plasma waves,'' {\em Journal of Russian Laser Research}, vol.~40,
  pp.~419--428, Sep 2019.

\bibitem{Arber2015}
T.~D. Arber, K.~Bennett, C.~S. Brady, A.~Lawrence-Douglas, M.~G. Ramsay, N.~J.
  Sircombe, P.~Gillies, R.~G. Evans, H.~Schmitz, A.~R. Bell, and C.~P. Ridgers,
  ``Contemporary particle-in-cell approach to laser-plasma modelling,'' {\em
  Plasma Physics and Controlled Fusion}, vol.~57, no.~11, p.~113001, 2015.

\bibitem{Fedele1986}
R.~Fedele, U.~de~Angelis, and T.~Katsouleas, ``Generation of radial fields in
  the beat-wave accelerator for gaussian pump profiles,'' {\em Physical Review
  A}, vol.~33, no.~6, pp.~4412--4414, 1986.

\bibitem{Gorbunov1996}
L.~Gorbunov, P.~Mora, and J.~T.~M. Antonsen, ``Magnetic field of a plasma wake
  driven by a laser pulse,'' {\em Physical Review Letters}, vol.~76, no.~14,
  pp.~2495--2498, 1996.

\bibitem{Gorbunov1997}
L.~M. Gorbunov, P.~Mora, and T.~M. Antonsen, ``Quasistatic magnetic field
  generated by a short laser pulse in an underdense plasma,'' {\em Physics of
  Plasmas}, vol.~4, no.~12, pp.~4358--4368, 1997.

\bibitem{Dawson1959}
J.~M. Dawson, ``Nonlinear electron oscillations in a cold plasma,'' {\em
  Physical Review}, vol.~113, pp.~383--387, Jan 1959.

\bibitem{Cowley2017}
J.~Cowley, C.~Thornton, C.~Arran, R.~J. Shalloo, L.~Corner, G.~Cheung, C.~D.
  Gregory, S.~P.~D. Mangles, N.~H. Matlis, D.~R. Symes, R.~Walczak, and S.~M.
  Hooker, ``Excitation and control of plasma wakefields by multiple laser
  pulses,'' {\em Phys. Rev. Lett.}, vol.~119, p.~044802, Jul 2017.

\bibitem{EpochGit}
``{EPOCH Particle-In-Cell code for plasma simulations}.''
  \url{https://github.com/epochpic/epochpic.github.io}.

\end{thebibliography}

\end{document}